\documentclass[aps,twocolumn,preprintnumbers,prl,amsmath,
amssymb,amsfonts,superscriptaddress,floatfix]{revtex4}
\usepackage{amsfonts}
\usepackage{amssymb}
\usepackage{amsmath}
\usepackage{ulem}
\usepackage{color}
\usepackage{latexsym}
\usepackage{graphicx}
\usepackage{graphics}
\usepackage{floatflt}
\usepackage{epsfig}
\usepackage{overpic}
\usepackage{soul} 
\usepackage[latin1]{inputenc}

\newcommand{\be}{\begin{equation}}
\newcommand{\ee}{\end{equation}}
\newcommand{\bea}{\begin{eqnarray}}
\newcommand{\eea}{\end{eqnarray}}

\def\lb{\label}

\def\G{\Gamma}
\def\D{\Delta}

\newcount\bozza \bozza=0
\ifnum\bozza=1
\newdimen\shift \shift=-2truecm
\def\lb#1{%
{\label{#1}\rlap{\kern\shift{$\scriptstyle#1$}}}}
\else\def\lb#1{\label{#1}} \fi

\begin{document}
\title{Synergy between Hund-driven correlations and boson-mediated Superconductivity}
\author{Laura Fanfarillo}
\affiliation{Scuola Internazionale Superiore di Studi Avanzati (SISSA) and 
CNR-IOM, Via Bonomea 265, 34136 Trieste, Italy}
\affiliation{Department of Physics, University of Florida, Gainesville, Florida 32611, USA}
\author{Angelo Valli}
\affiliation{Scuola Internazionale Superiore di Studi Avanzati (SISSA) and 
CNR-IOM, Via Bonomea 265, 34136 Trieste, Italy}
\affiliation{Institute for Solid State Physics, Vienna University of Technology, 1040 Vienna, Austria}
\author{Massimo Capone}
\affiliation{Scuola Internazionale Superiore di Studi Avanzati (SISSA) and 
CNR-IOM, Via Bonomea 265, 34136 Trieste, Italy}

\begin{abstract}{Multiorbital systems such as the iron-based superconductors provide a new avenue to attack the longstanding problem of superconductivity in strongly correlated systems. In this work we study the superconductivity driven by a generic bosonic mechanism in a multiorbital model including  the full dynamical electronic correlations induced by the Hubbard $U$ and the Hund's coupling. We show that superconductivity survives much more in a Hund's metal than in an ordinary correlated metal with the same degree of correlation. The redistribution of spectral weight characteristic of the Hund's metal reflects also in the enhancement of the orbital-selective character of the superconducting gaps, in agreement with experiments in iron-based superconductors.}

\end{abstract}

{\bf \maketitle }
Thirty years of research have established a new paradigm in which strong electronic correlations and Mott physics are strongly intertwined with superconductivity (SC). 
A popular point of view, developed mainly for the copper-based superconductors,  requires to abandon the main concepts of the BCS theory, including the very idea that pairing is mediated by a bosonic glue \cite{Anderson_Science1987,Lee_RMP2006}. For other classes of materials, including the iron-based superconductors (IBS), it seems more appropriate to adopt an intermediate picture where a boson-mediated SC coexists or even benefits from the presence of strong electron-electron interactions.
An important step in this direction has been taken in \cite{Capone_Science2002}, where it has been shown that the proximity to a Mott transition can strongly boost phonon-driven superconductivity as long as the phonons do not couple with charge fluctuations, which are frozen in the Mott insulator \cite{Han_PRL2003}. The idea has been discussed in a model for alkali-doped fulleride \cite{Capone_RMP,Nomura}, and it has also been connected with the physics of cuprates \cite{Capone_PRL2004,Schiro}. In this perspective, the antiferromagnetic superexchange in the copper-oxygen layers  plays the role of the built-in pairing mechanism which survives to Mott localization as it involves the spin degree of freedom. 

IBS are an important piece of this puzzle, as they feature non-trivial electron-electron correlations effects \cite{DeMedici_Chapter2015} but, at the same time SC can be successfully described in terms of itinerant electrons coupled by the exchange of bosons \cite{Mazin_PRL2008, Kuroki_PRL2008, Kontani_PRL2010, Chubukov_Chapter2015}. A number of theoretical and experimental works have clarified that SC in IBS emerges from a bad metallic phase characterized by a multiorbital electronic structure and a sizeable value of the Hund's coupling \cite{Yin_NatMat2011, Werner_NatPhys2012, Hardy_PRL2013, DeMedici_PRL2014, Maletz_PRB2014, Yi_NatComm2015, McNally_PRB2015, Backes_PRB2015, Hardy_PRB2016, Lafuerza_PRB2017, Watson_PRB2017}. This Hund's metal is a incoherent metallic state where Hund's driven correlations lead to low coherence temperature \cite{Georges_Review2013, Werner_PRL2008, Haule_NJP2009, DeMedici_PRL2009, Ishida_PRB2010, Liebsch_PRB2010, Yin_NatMat2011, deMedici_PRB2011, DeMedici_PRL2011, Yu_PRB2012, Bascones_PRB2012, Lanata_PRB2013, DeMedici_PRL2014, Fanfarillo_PRB2015, Stadler_AP2018, Isidori_PRL2019, Mezio_arxiv2019} and the correlation effects are strongly orbital selective: electrons occupying different orbitals can have substantially different effective mass and scattering rate \cite{DeMedici_PRL2014, DeMedici_Chapter2015, Capone_NatMat2018, Mezio_arxiv2019}.

The evidence of strong correlation may appear as a challenge to the claim that a theory based on itinerant electrons can explain the superconducting properties of these materials. In that respect, it has been recently shown that in many cases phenomenological weak-coupling approaches are able to describe the experimental scenario only if combined with orbital-dependent properties descending from electronic correlations. A notable example is the description of the FeSe anisotropic gap functions in the Brillouin zone \cite{Sprau_Science2017, Kushnirenko_PRB2018, Rhodes_PRB2018} in terms of phenomenological approaches based either on an orbital-dependent single-particle renormalization \cite{Kreisel_PRB2017, Hu_PRB2018, Kreisel_PRB2018} or orbital-selective pairing \cite{Benfatto_npjQM2018}.

We believe that such coexistence of strong correlation physics and properties associated with boson exchange calls for an understanding of SC in IBS where the two phenomena are treated on equal footing.  We attack this problem with an approach inspired by Ref.~\cite{Capone_Science2002}. We assume that SC is driven by some kind of weak-coupling mechanism (e.g. the coupling to a boson), while the normal state contains the dynamical correlations of the Hund's metal.
For the sake of definiteness, we consider a simplified multiorbital model for IBS including an explicit pairing while we account for electronic correlations dressing the particle-particle propagator with orbital and frequency-dependent self-energies which include all the properties of a Hund's metal as described by Dynamical Mean-Field Theory (DMFT). As a matter of fact we study a superconductor where the Cooper pairs are formed by fully dressed electrons. 
Our main finding is that the dynamical properties of the Hund's metal positively affect the superconducting instability, in fact SC survives in a Hund's metal much more than in an ordinary correlated metal characterized by the same effective mass renormalization and same density of states at the Fermi level. The redistribution of spectral weight in the Hund's metal also enhances the orbital-selective character of the superconducting gaps in agreement with recent experiments in IBS \cite{Sprau_Science2017}.

%
\begin{figure}
\includegraphics[width=0.96\linewidth]{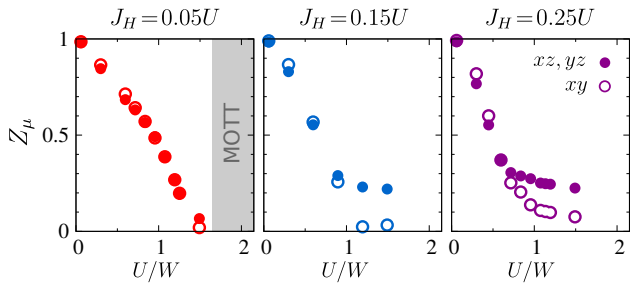}
\caption{Quasiparticle weights $Z_{\mu}$ as a function of $U$ for different values of $J_H$. $U$ is in unit of the bare bandwidth $W\sim 1.6$~eV. For $J_H=0.05~U$ the $Z$ are suppressed monotonously as $U$ increases and the Mott phase is established for $U\sim1.6~W$. At larger $J_H/U$ values the Mott phase is pushed towards higher values of $U$ and one recovers the typical features of the Hund's metal phenomenology.}
\vspace{-.6cm}
\label{fig:DMFT_Z}
\end{figure}

We consider a minimal model which accounts for the main features of the electronic structure of IBS and for the electron-electron correlations induced by the combined effect of the Hubbard repulsion and the Hund's coupling \cite{suppl}. The kinetic Hamiltonian is given by a three-orbital tight-binding model adapted from \cite{Daghofer_PRB2010}. This reproduces qualitatively the shape and the orbital content of the Fermi surfaces typical  of the IBS family, namely two hole-like pockets composed by $yz$-$xz$ orbitals around the $\G$ point and two elliptical electron-like pockets formed by $xy$ and $yz/xz$ orbitals centered at the $X/Y$ point of the 1Fe-Brillouin Zone. Local electronic interactions are included considering the multiorbital Kanamori Hamiltonian which parametrizes the electron-electron interactions in term of a Hubbard-like repulsion $U$ and an exchange coupling $J_H$ favoring high-spin states \cite{Georges_Review2013}.

The effect of the interactions is contained within DMFT in the  ${\bf k}$-independent self-energy $\Sigma_{\mu\mu}(i \omega_n)$, where $\mu$ is the orbital index and $\omega_n$ is the $n$-th fermionic Matsubara frequency. We solve DMFT using an exact diagonalization solver at zero temperature \cite{Capone_PRB2007, Weber_PRB2012} for fixed $U$ and $J_H$ at a density of four electrons in three orbitals per site, that, while reproducing the  low-energy electronic structure with hole and electron pockets, also gives rise to the Hund's metal features, analogously to the filling of six electrons in five orbitals characteristic of IBS. 

In order to highlight the role of the frequency dependence of the self-energy, we will compare the full DMFT results with an approximate quasiparticle (QP) picture in which the effects of the interaction are encoded via the QP spectral weight $Z_{\mu}= (1 - \partial \Im \Sigma_{\mu\mu}/\partial \omega_n)^{-1}$ and an energy shift $\D \Sigma_{\mu} = lim_{\omega_n \to 0} \Re\Sigma_{\mu\mu}(i\omega_n)$\footnote{In all the DMFT calculations we find $\Im\Sigma_{\mu\mu}(0) =0$ at zero temperature.}. The orbital-dependent QP weight $Z_{\mu}$ measures the correlation-induced reduction of the coherent behavior of the electrons, and within DMFT, coincides with the inverse of the effective mass enhancement.

We consider SC in the intraorbital spin-singlet channel only, which implies an orbital-diagonal gap function $\Delta_{\mu}$ and a pairing Hamiltonian of the form $H_{SC} =  - \sum_{\mu \nu} g_{\mu\nu} {\Delta_{\mu}}^{\dagger} \Delta_{\nu}$, where $g_{\mu \nu}$ is the superconducting coupling that allows pair hopping from the $\mu$ orbital to the $\nu$. We restrict to intraorbital pair hopping, $g_{\mu \mu}\!=\!g$ for each orbital $\mu=yx,xz,xy$. 
Such diagonal coupling in the orbital basis reflects itself into inter-pocket components of pairing in the band representation in agreement with theoretical modeling in IBS \cite{Chubukov_Chapter2015}. We fix $g=2$~eV, which leads, in the absence of interactions, to gaps around half of the bandwidth $W$, a sizeable values which makes the numerical analysis simple without spoiling its general character. In what follows we show the outcomes of numerical calculations that assume $g$ to be an energy-independent coupling. While boson-mediated pairing interaction takes large value only around the Fermi energy, in the SM \cite{suppl} we show that our conclusions are robust and independent on the choice of a cut-off.

In Fig.~\ref{fig:DMFT_Z} we plot $Z_{\mu}$  as a function of $U$ for three representative values of $J_H$. We recover the typical features of the Hund's metal crossover. While at $J_H=0.05~U$ the QP spectral weights are suppressed monotonously as $U$ increases, for larger $J_H$ values a faster initial suppression of the $Z$ is followed by a long tail \cite{DeMedici_PRL2011} after a crossover located around $U\lesssim W$ where a differentiation between the $xz/yz$ and the $xy$ orbitals sets in \cite{DeMedici_PRL2014}. 

\begin{figure*}
\centering
\includegraphics[clip,width=0.98\textwidth]{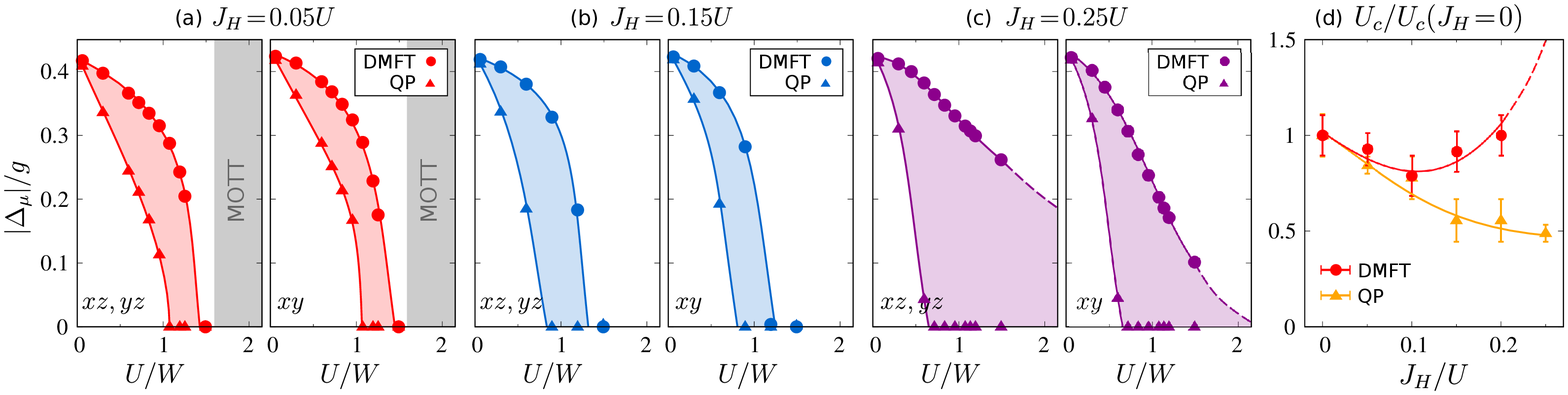}
\caption{(a-c) BCS solutions for the orbital gaps as a function of $U$ for $J_H/U = 0.05, 0.15, 0.25$ and $g =2$ eV. Larger values of $J_H/U$ boost the SC up to very high values of $U$. Colored regions emphasize the difference between the DMFT and the QP results. (d) Critical $U$ at which the gaps close as a function of $J_H$, renormalized to the $J_H=0$ value. $U_c$ increases with $J_H$ within the full DMFT calculation while decreases within the QP approximation following the $Z$ behaviour.} 
\vspace{-.5cm}
\label{fig:GapJh}
\end{figure*}

In Fig.~\ref{fig:GapJh} we show the orbital-resolved zero-temperature superconducting gaps $\Delta_{yz}=\Delta_{xz}$ and $\Delta_{xy}$ for the same parameters of Fig.~\ref{fig:DMFT_Z}. We solve the standard BCS gap equation where the Cooper bubbles are computed using the fully-dressed Green's functions obtained by DMFT. As mentioned above, we also consider the QP approximation introduced above \cite{suppl}.
At small $J_H$, Fig.~\ref{fig:GapJh}a,  the gaps are reduced by increasing $U$ and they vanish at a critical strength $U_c \sim W$. The full DMFT self-energy results provide larger gaps than the QP analysis. This signals that the dynamical effects beyond a simple QP picture reduce the negative effect of repulsive interactions, even if they do not critically affect $U_c$. 
Upon increasing $J_H$, Fig.~\ref{fig:GapJh}b, \ref{fig:GapJh}c, the difference between the QP approximation and DMFT becomes striking and we observe both a significant difference in the gaps and a remarkable enhancement of the critical repulsion needed to destroy the superconducting phase. We find therefore a large window of parameters where SC survives in the presence of strong interactions despite very small values of the QP weights $Z$ which inevitably kill the order parameter in a simple QP approximation. 

A plot of the critical interaction $U_c$ as a function of the $J_H$ ratio reveals that $J_H$ does not simply reduce the negative effect of repulsion on SC, but it leads to a significant increase of the superconducting region in the phase diagram, in sharp contrast with the result of the QP approximation which shows a reduction of $U_c$ with $J_H$, see Fig.~\ref{fig:GapJh}d. Therefore the Hund's metal emerges as a more favourable environment to develop SC with respect to a more standard strongly renormalized Fermi-liquid described by the QP approximation.

The above results indicate that at small $J_H$ the low-energy features of $\Sigma_{\mu\mu}(\omega)$, encoded in the QP approximations, retain the most important physical information, in the Hund's metal regime instead the spectral weight at finite energy scales strongly affect the particle-particle propagator. In order to put this observation on solid ground, we discuss the frequency dependence of the normal-state  spectral function $A({\bf k},\omega)$.

\begin{figure}[b]
\centering
\includegraphics[width=0.96\linewidth]{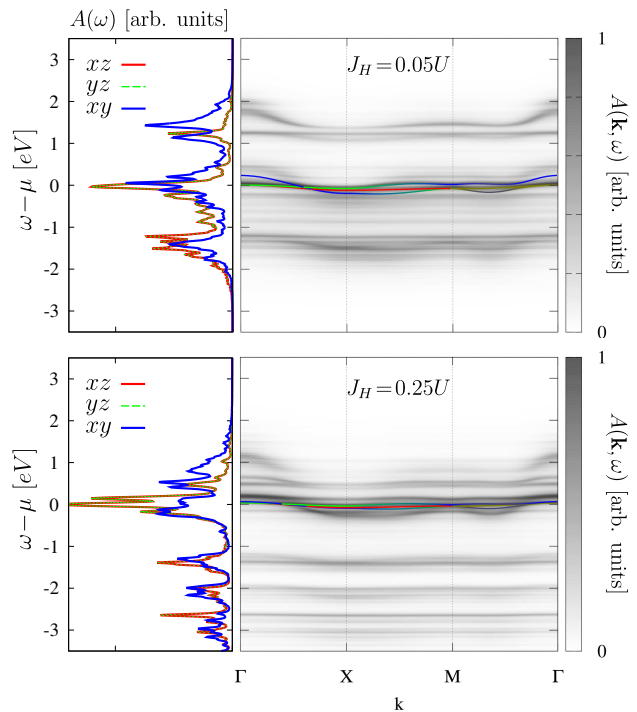}
\vspace{-.2cm}
\caption{Local DOS and ${\bf k}$-resolved spectral function $A({\bf  k},\omega)$ at $U \sim 1.2~W$ and $J_H/U = 0.05, 0.25$. Hubbard bands develop at an energy scale of order $U$ for $J_H/U=0.05$. For $J_H/U=0.25$ the spectral weight is redistributed over a window of energy of order $J_H$ around the Fermi level. The color lines superimposed to $A({\bf  k},\omega)$ show the strongly renormalized bands in the QP approximation.} 
\label{fig:Akw}
\end{figure}

In Fig.~\ref{fig:Akw} we compare the spectral weight redistribution in two correlated regimes having similar values of $Z_{\mu}$ (cfr. Fig.~\ref{fig:DMFT_Z}) but characterized by different values of the Hund's coupling: $J_H = 0.05~U$, for which the DMFT and QP analysis of the SC gives similar results and $J_H=0.25~U$, inside the Hund's metal regime, where the QP approximation is not able to capture the boost of SC due to $J_H$.  
The two spectral functions  have a similar effective bandwidth for the low-energy QP excitations as a consequence of the similar $Z_{\mu}$.
However, the two regimes show remarkable differences in the higher-energy excitations which live on different scales. For $J_H=0.05~U$ we find a Mott-like behavior where Hubbard bands develop at energy scales of order $U$, while for $J_H=0.25~U$ the redistribution of spectral weight mainly accumulates in a narrow energy window of order $J_H$ around the Fermi level as also discussed in \cite{deMedici_PRB2011, Werner_NatPhys2012, Backes_PRB2015, Stadler_AP2018}.
This picture explains the results for SC of Fig. \ref{fig:GapJh}. At small $J_H/U$, when $U$ is the dominant energy scale, high- and low-energy features are largely decoupled. Only the low-energy features of the $\Sigma_{\mu\mu}(\omega)$ affect the excitations close to the Fermi energy that are relevant to determine the pairing instability and the superconducting gaps.
In the Hund's metal regime instead the spectral weight redistribution mainly occurs in a range of order $J_H$ around the Fermi energy. This finite-frequency contributions encoded in the DMFT self-energy contribute to the pairing together with the QP contributions, thereby enhancing the superconducting tendency with respect to the QP approximation. The analysis performed using a finite cut-off energy in the pairing interaction further confirms the involvement of the low-energy spectral weight in SC \cite{suppl}. Let us finally stress again that both the low and high $J_H$ regimes shown here are characterized by similar $Z$ factors, thus the different behavior of the gaps function can not be interpreted by the Fermi-liquid renormalization of the density of states that scales as $1/Z$, and have instead to be ascribed to the different spectral weight redistribution of the Hund's metal. 
 
\begin{figure}[t]
\centering
\includegraphics[width=\linewidth]{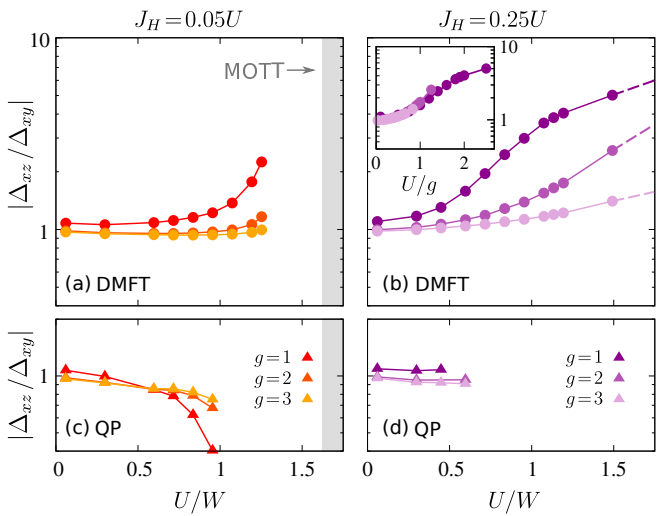}
\vspace{-.2cm}
\caption{$\Delta_{xz}/\Delta_{xy}$ as a function of $U$ for $J_H/U = 0.05$ and $0.25$, and $g=1,2,3$ eV within (a-b) the full DMFT self-energy dressed superconducting bubble calculation, (c-d) the QP approximation. Strong orbital gap differentiation is found in the Hund's metal regime for the DMFT calculation only. Inset: data collapse of the main panel as a function of $U/g$}
\label{fig:Ratiogap}
\vspace{-.5cm}
\end{figure}

The orbital character of the superconducting gaps shown in Fig.~\ref{fig:GapJh} can be rationalized by plotting the ratio $|\D_{yz/xz}|/|\Delta_{xy}|$. The orbital-selective character of the electronic properties is indeed a distintive feature of Hund's metals \cite{Capone_NatMat2018} which has been discussed in relation with recent experiments for FeSe \cite{Sprau_Science2017,Kushnirenko_PRB2018, Rhodes_PRB2018} and 122 compounds \cite{Evtushinsky_PRB2014, Hardy_PRB2016}.
Fig.~\ref{fig:Ratiogap} shows $|\D_{xz}|/|\Delta_{xy}|$ as a function of the interaction parameters and of the superconducting coupling. From the DMFT results, Fig.~\ref{fig:Ratiogap}a, \ref{fig:Ratiogap}b, we find that for $J_H=0.05~U$ the ratio is constant and $\sim 1$ up to $U\lesssim U_c$. On the other hand, for $J_H=0.25~U$ the orbital differentiation grows monotonically with $U$ and is already large at small value of $U$. As demonstrated by the data collapse in the inset, the different values of $U$ at which the differentiation starts to grow only depends on the distance from the critical interaction $U_c$ which in turn increases as the coupling $g$ is larger. 

Fig.~\ref{fig:Ratiogap}c, \ref{fig:Ratiogap}d show that the strong orbital selectivity of the superconducting gaps in the Hund's metal is not captured by a QP approximation. In this case in fact the gaps ratio is $\sim 1$ up to the gaps closure both in the low- and high- $J_H$ regimes. This result can be connected with the QP analysis of orbital-selective SC in FeSe \cite{Sprau_Science2017, Kreisel_PRB2017, Kreisel_PRB2018}, where strongly orbital-selective $Z$'s with surprisingly large differentiations were required to reproduce the experimental results. Similar values of the QP renormalization are hardly predicted by solutions of models including multiorbital local interactions. Based on our analysis, we can view  the extreme estimates of $Z$ of \cite{Sprau_Science2017, Kreisel_PRB2017, Kreisel_PRB2018} as the result of effectively including the frequency-dependence of the Hund's metal self-energy in a single parameter. 
%

In this work we assumed an orbital-diagonal superconducting coupling as an unbiased model choice to single out the role of non-trivial dynamical correlations characteristic of the Hund's coupling to stablize SC. 
We do not expect that specific and more realistic and material-dependent choices of the coupling can change the qualitative results as long as the main ingredients of our toy model are preserved. In particular, in our analysis we did not introduce any cut-off energy for the superconducting interaction, i.e. we implicitly assumed that the frequency range of the superconducting coupling is comparable or larger than the energy scale over which the Hund's metal spectral weight is redistributed. This is expected to be realized in IBS due to the large value of the Hund's coupling and the quite large frequency range of the spin/orbital mediated superconducting interaction. In the SM we provide a detailed analysis in terms of the cut-off energy that shows the validity of our conclusions \cite{suppl}. In our calculations we further assumed that the superconducting channel is not strongly renormalized by the Coulomb repulsion.

This idea is inspired by a series of results for multiorbital models for alkali-metal doped fullerides \cite{Capone_Science2002}. The crucial aspect is that the phonon-driven attraction has the form of an inverted Hund's coupling, i.e. it only involves local spin and orbital degrees of freedom. As a consequence the superconducting coupling is not renormalized by a large $U$, which freezes only charge fluctuations but leaves the local spin and orbital channels free to fluctuate. Thus, when the system approaches the Mott transition, the heavy QP with small $Z$ experience an effective pairing interaction of the same order of the bare one.
We believe that the same idea holds for the pairing mechanisms relevant for IBS or other multicomponent strongly correlated superconductors in which non-local spin- or orbital-fluctuations act as mediators for the SC.

The same non local character of the fluctuations mediating pairing is the main theoretical difficulty in directly extending the scenario of \cite{Capone_Science2002} to  IBS. 
However, the main features we identified within the present simplified approach are expected to survive in more complete and realistic description of the the IBS based for example on cluster or diagrammatic extensions of DMFT for multiorbital models.
In this sense our work can be seen as complementary to the analysis of spin-mediated pairing of \cite{Nourafkan_PRL2016}, where the superconducting vertex mediated by magnetic excitations is computed within the random phase approximation on a correlated electronic structure. In \cite{Nourafkan_PRL2016} the authors focus on the analysis of the symmetry of the order parameter induced by correlation effects, however they do not explicitly investigate the role of $J_H$ as booster of the spin-mediated pairing.


In conclusion, we have studied how a pairing based on the exchange of bosons (phonons or spin/orbital fluctuations) coexists with the electronic correlations induced by the Hund's coupling.
Solving a three-orbital model inspired by the electronic structure of the IBS we study the pairing instability assuming that single-particle properties are renormalized nonperturbatively by the interactions $U$ and $J_H$. 
The main result is that, for a Hund's metal, the loss of spectral weights at the Fermi level (measured by the QP weight $Z$) does not imply the suppression of SC once non-trivial dynamical correlations are taken into account. This is a consequence of a spectral weight redistribution which does not follow a standard Mott-like behavior but leads to the population of states in an energy window of order $J_H$ around the Fermi level.
Dynamical correlations also crucially affect the orbital-selective nature of the superconducting gaps. Despite the small orbital differentiation in the QP weights, the finite-frequency correlations enhance the orbital differentiation of the gaps in the Hund's metal regime explaining why previous QP analysis of FeSe required to introduce hugely orbital selective $Z$'s. 

We conclude pointing out that the low-energy spectral-weight redistribution in Hund's metals can in principle enhance also instabilities in the particle-hole channel, e.g nematicity. Preliminary results suggest that the inclusion of finite frequency correlation effects could indeed affect the picture previously obtained within QP approximation analysis \cite{Fanfarillo_PRB2017}. However, in contrast with SC, the nematic order introduces not trivial modifications of the band structure that makes the analysis extremely sensitive to the parameter choice. A thorough study is beyond the scope of the manuscript and will be properly investigated in a future work.

\section*{ACKNOWLEDGEMENTS}
We are grateful to C.~Castellani and L.~de'~Medici for helpful discussions. L.~F. acknowledges financial support from the European Unions Horizon 2020 research and innovation programme under the Marie Sklodowska-Curie grant SuperCoop (Grant No 838526). A.~V. acknowledges financial support from the Austrian Science Fund (FWF) through the Erwin Schr\"{o}dinger fellowship J3890-N36. M.C. acknowledges  financial  support from  MIUR  PRIN  2015  (Prot. 2015C5SEJJ001)  and SISSA/CNR project ``Superconductivity, Ferroelectricity and Magnetism in bad metals'' (Prot.  232/2015).

\bibliography{Iron_SC}

\end{document}